\documentclass[a4paper,11pt]{article}
\usepackage{a4wide,cite}

\makeatletter
\gdef\@publabel{\hfil}
\gdef\@pubnumber{\null}
\gdef\@author{\null}
\gdef\@title{\null}
\gdef\@abstract{\null}
\long\def\pubnumber#1{\gdef\@pubnumber{#1}}
\long\def\publabel#1{\gdef\@publabel{#1}}
\long\def\author#1{\gdef\@author{#1}}
\long\def\title#1{\gdef\@title{#1}}
\long\def\abstract#1{\gdef\@abstract{#1}}
\def\titlerelax{
\let\maketitle\relax
\let\settitleparameters\relax
\let\consolidatetitle\relax
\let\inittitlepage\relax
\let\finishtitlepage\relax
\let\titlepagecontents\relax
\let\multithanks\relax
\let\titlebaselines\relax
\let\@makepub\relax
\let\@maketitle\relax
\let\@makeauthor\relax
\let\@makeabstract\relax
\let\@maketitlenote\relax
\let\thanks\relax
\let\titlerelax\relax}
\def\titleclean
{\gdef\@titlenote{}
\gdef\@abstract{}
\gdef\@author{}
\gdef\@title{}
\gdef\@dpublabel{}
}

\def\@makepub{\vbox to \z@{\hbox to \textwidth{\hfill
\@publabel \hfill
\llap{\parbox[t]{0.33\textwidth}{\raggedleft\@pubnumber}}}%
\vss}}
\def\@maketitle{\vskip 60pt \begin{center}
 {\LARGE \@title \par}
 \end{center}}

\def\@makeauthor{{%
\def\and{\smallskip {\normalsize \rm and\smallskip }}
\def\And{\medskip {\normalsize \rm and\\}\medskip}
\long\def\address##1{{\def\and{\\and\\}\medskip
                                {\small \it \\##1\\}
}}
{\centering
 \vskip 3em
 \large \lineskip .75em
 \@author}
 \par}}
\def\@makeabstract{\vskip 1.5em
{\small
\begin{center}
{\bf ABSTRACT\vspace{-.5em}\vspace{0pt}}
\end{center}
\quotation \@abstract \endquotation}}
\def\maketitle{\titlepage
\let\footnotesize\small \setcounter{page}{0}
\@makepub
\vfil
\@maketitle
\@makeauthor
\vfil
\@makeabstract
\@thanks
\vfil
\titlerelax \titleclean
\setcounter{footnote}{0}
}
\makeatother

\def\bea{\begin{eqnarray}}
\def\be{\begin{equation}}

\def\cnm1{{$ \smash{ c^{(1)}_{\scriptscriptstyle{(n-1)/2}}}$}}

\def\d{{\mathrm d}}
\def\eea{\end{eqnarray}}
\def\ee{\end{equation}}

\def\nn{\nonumber\\}

\def\qptry2(#1){T_{#1}}

\def\U {\tilde U}

\def\vec#1{{ \left| #1 \right\rangle}}
\def\vac[#1]{ \left( #1 \right) \vec 0}

\def\WB{{W\!B}}

\begin{document}

\parindent 0pt
\parskip 5pt

\title{Conserved charges in the chiral 3-state Potts model}
\author{G.M.T. Watts
\address{Department of
Mathematics, King's College London, The Strand, London, WC2R 2LS}
}
\pubnumber{KCL-MTH-97-51\\
hep-th/9708167\\
August 29, 1997}

\abstract{%
We consider the perturbations of the 3-state Potts conformal field
theory introduced by Cardy as a description of the chiral 3-state
Potts model. By generalising Zamolodchikov's counting argument and by
explicit calculation we find new inhomogeneous conserved currents for
this theory. We conjecture the existence of an infinite set of
conserved currents of this form and discuss their relevance to the
description of the chiral Potts models.
}

\maketitle
\section{Introduction}

The chiral 3-state spin chain (see e.g. \cite{HKDN1,ints}) has
Hamiltonian
\be
  H = - \frac{2}{\sqrt 3}\, \sum_j \;
  \left( e^{-i\varphi/3} \sigma_j^{} + e^{i\varphi/3} \sigma_j^\dagger \right)
+ \lambda \left(
   e^{-i\phi/3} \Gamma_j^{} \Gamma_{j+1}^\dagger +
   e^{i\phi/3}  \Gamma_j^\dagger \Gamma_{j+1}^{} \right)
\;,
\label{eq:sch}
\ee
where $j$ labels the sites and the matrices $\sigma_j$ and $\Gamma_j$
at each site are
\[
  \sigma = \pmatrix{ 1 & 0 & 0 \cr 0 & \omega & 0 \cr 0 & 0 & \omega^2}
\;,\;\;
  \Gamma = \pmatrix{ 0 & 0 & 1 \cr 1 & 0 & 0 \cr 0 & 1 & 0 }
\;,
\]
with $\omega = \exp( 2 \pi i /3 )$.
If $\cos\varphi = \lambda\cos\phi$ then the model is known to be
integrable \cite{ints}, and is self-dual for $\phi =\varphi$. The
`standard' 3-state Potts model is obtained by setting
$\varphi=\phi=0$, and this has a second order phase transition at
$\lambda=1$ which is described by a conformal field theory
with $c=4/5$ \cite{Dots2}.

The standard model can be viewed as a perturbation of this conformal
field theory by a particular field \cite{Dots2}, known as the thermal
perturbation ($\lambda$ corresponding to temperature),
and it is known that the resulting massive field theory is integrable
with an infinite set of holomorphic and anti-holomorphic conserved
quantities \cite{Zamo4}.
The existence of several of these conserved quantities can be shown by
a simple counting argument due to Zamolodchikov \cite{Zamo4,Zamo2b}.
It is interesting to note here that the corresponding perturbation of
the lattice model is probably not integrable in the usual sense.

In \cite{Card3}, Cardy considered the possibility that
the full chiral Potts model can be viewed as a perturbation of
this conformal field theory. He pointed out that half of the conserved
quantities (the anti-holomorphic, say) of the thermal perturbation
remain conserved when this an extra, chiral, perturbation is
added and identified this doubly perturbed model depending on two free
parameters with the self-dual sub-sector of the general chiral 3-state
Potts model, but the relation between the integrable sub-sector of the
chiral spin chain and this integrable field theory was unclear.
However, it is still remarkable that such a perturbation by two
relevant fields is integrable as the only double perturbations
previously known to be integrable are the staircase models
\cite{stair} which have one relevant and one irrelevant perturbation.

One question was whether there are any counterparts in the
doubly perturbed theory to the holomorphic conserved quantities of the
thermal perturbation. In this article we show that the double
perturbations does have further conserved quantities which reduce to
the `missing' holomorphic conserved quantities when the `extra'
perturbation is removed, and conjecture that all the conserved
quantities of the thermal perturbation can be extended in this way to
conserved quantities of the doubly perturbed model.

Zamolodchikov's counting argument generalises to the
case of a double perturbation and proves the existence of two extra
conserved quantities; we have checked explicitly the
existence of two more, but at the moment a general proof is lacking.

The outline of this letter is as follows.
We first recall the conformal field theory treatment of the Potts
model, the conserved quantities for the various perturbations, and how
the existence of several of these can be deduced by Zamolodchikov's
counting argument.

We then consider the conserved quantities of the doubly perturbed
model, give a counting argument and a few explicit examples of
conserved quantities of this model.

Finally we speculate on the possible implications of these results for
the doubly-perturbed and general perturbations of the conformal Potts
model and the relation of the perturbed conformal field theories to
the spin chain.

\section{The conformal field theory of the 3-state Potts model}

The 3-state Potts model was one of the first conformal field theories
to be studied \cite{Dots2} and is both a minimal
model of the Virasoro algebra and of the $W_3$ algebra \cite{Zamo1},
and hence the same field content admits two descriptions.

The Potts model has $c=4/5$ and has Virasoro primary fields of weights
\bea
  (h,\bar h) &=&
  (0,0),\,(3,0),\,(0,3),\,(3,3),\,(2/3,2/3),\,(1/15,1/15),
\nn&&
  (2/5,2/5),\,(2/5,7/5),\,(7/5,2/5),\,(7/5,7/5)
\;.
\eea
The field with weights $(3,0)$ is the holomorphic generator $W(z)$ of
the $W_3$ algebra, and we take its commutation relations to be
\bea
    ~[\,W_m\,,\,W_n\,]
&=&
    {\displaystyle \frac {13}{10800}}\,m(m^2-1)(m^2-4) \delta_{m+n}
 \\
&& + {\displaystyle \frac {13}{720}}\,(\,m-n\,)\,
(\,2\,m^2 - mn + 2\,n^2 - 8\,)\,L_{m+n}
  + {\displaystyle \frac {1}{3}} \,(\,m-n\,)\, \Lambda_{m+n}
\;.
\nonumber \eea
By allowing non-standard normalisations for the fields
$(2/5,7/5),(7/5,2/5),(7/5,7/5)$ we can identify them as
descendants of the $W$-primary field $(2/5,2/5)$ as follows:
\[
  \vec{7/5,2/5} = W_{-1} \vec{2/5,2/5}
\;,\;\;
  \vec{2/5,7/5} = \bar W_{-1} \vec{2/5,2/5}
\;,\;\;
  \vec{7/5,7/5} = W_{-1} \bar W_{-1} \vec{2/5,2/5}
\;.
\]

\subsection{Integrable perturbations and conserved quantities}

The most general action for a perturbed conformal theory is
\[
  S = S_0 + V
\;,\;\;
 V = \int\! \d^2 z\; \Phi(z,\bar z)
\;,
\]
where $S_0$ is the action of the conformal field theory and the
potential $V$ is given in terms of some local field $\Phi(z,\bar z)$.

A field $U(z)$ which is holomorphic in the unperturbed theory may
develop $\bar z$ dependence in the perturbed theory since its
correlation functions are given as
\[
  \langle  U(z,\bar z) \ldots \rangle
= \sum_n {1\over{n!}}
  \langle
  U(z) V^n \ldots
  \rangle_0
\;,
\]
where $\langle\;\rangle_0$ is calculated in the unperturbed theory.
To first order,
\be
  \bar\partial\,U(z,\bar z)\, =\, \chi(z,\bar z)
\;,
\label{eq:fo}
\ee
 where $\chi$ is the residue of the simple pole in the operator
product expansion of $U$ and $v$,
\be
  U(z) \; \Phi(w,\bar w) \; = \; \ldots + \frac{ \chi(w,\bar w) }{z - w}
  + \ldots
\;.
\label{eq:ope}
\ee
If $\chi(w,\bar w)$ is a total $z$--derivative, i.e. if $\vec\chi =
L_{-1}\vec\xi$ for some state $\vec\xi$, then  $\oint\!\d z\; U$
remains conserved to first order in perturbation theory. By an abuse
of notation we shall say that $U$ is conserved if $\bar\partial U =
\partial\xi$ for some $\xi$. It is sometimes possible to show that no
higher order corrections are possible and that in this way a quantity
is conserved to all orders in the perturbed theory.
The $\bar w$ dependence of $\Phi$, $\chi$ and $\xi$ is essentially
irrelevant when checking the conservation of $\oint\!\d z\; U(z)$ to
first order, and so we drop this dependence for the rest of this
section. We shall keep the notation $\Phi_{h,\bar h}$ for fields with
$z$ and $\bar z$ dependence, and use $\phi_h$ or $\bar\phi_{\bar h}$
for the chiral dependence of such a field.

A large class of integrable perturbations of conformal field theories
are affine Toda field theories, for which the existence of an infinite
number of conserved quantities has been proven by Feigin and Frenkel
\cite{FFre7}. Two perturbations of the Potts model can be interpreted as
affine Toda field theories (ATFTs), namely the perturbations by
the fields $\phi_{2/5}$ and $\phi_{7/5}$. (n.b.\ the
$\Phi_{7/5,7/5}$ perturbation is irrelevant, and consequently there
may be higher order corrections to the conservation equation, but we
shall ignore the anti-holomorphic dependence of perturbations and such
issues.)

The $\phi_{7/5}$ perturbation corresponds to the $a_1^{(1)}$ ATFT, or
Sine-Gordon, theory which has conserved currents $\U_\Delta(z)$ of
weights $\Delta=2n$.

The $\phi_{2/5}$ perturbation corresponds both to the $a_2^{(2)}$ ATFT
with conserved currents $U_\Delta$ polynomial in $L$ of weights
$\Delta=6n,6n+2$, and also to the $a_2^{(1)}$ theory with conserved
currents polynomial in both $L$ and $W$ of weights $\Delta=3n,3n+2$.
In this case the currents of even weight $6n,6n+2$ are common to both
theories and are independent of $W$ whereas the currents of odd weight
$6n+3,6n+5$ are odd under $W \to -W$.

We give the first few states $U_\Delta = U_\Delta(0)\vec 0$ and
$\U_\Delta = \U_\Delta(0)\vec 0$ below. (n.b.\ these expressions are
not unique, since they are only defined up to the addition of total
derivatives and null states. We have chosen representatives which are
reasonably compact).

\begin{table}[htb]
\[
{\renewcommand{\arraystretch}{1.4}
\begin{array}{lll}
\hline
U_2 &=&  L_{-2}\,\vec 0
 \\
U_3 &=&  W_{-3}\,\vec 0
 \\
U_5 &=&  L_{-2} W_{-3} \,\vec 0
 \\
U_6 &=& \vac[ L_{-2} L_{-2} L_{-2}  +
         { \frac {21}{10}}\,L_{-3} L_{-3} ]
 \\
U_8 &=& \vac[ L_{-2} L_{-2} L_{-2} L_{-2}
      - { \frac {159}{25}}\,L_{-3} L_{-3} L_{-2}
      - { \frac {249}{25}}\,L_{-4} L_{-2} L_{-2} ]
 \\
U_9 &=& \vac[ L_{-2} L_{-2} L_{-2} W_{-3}
      + \frac{1071}{10}\, L_{-2}L_{-2}W_{-5}
      + \frac{16731}{125}\,L_{-3}W_{-6} ]
 \\[1mm] \hline
\U_2 &=& L_{-2} \, \vec 0
 \\
\U_4 &=& L_{-2} L_{-2}  \, \vec 0
 \\
\U_6 &=& \vac[ L_{-2} L_{-2} L_{-2}
  - \,{ \frac {7}{30}}\, L_{-3} L_{-3}  ]
 \\
\U_8 &=& \vac[ L_{-2} L_{-2} L_{-2} L_{-2}
   - { \frac {229}{375}}\, L_{-3} L_{-3} L_{-2}
   + { \frac {871}{375}}\, L_{-4} L_{-2} L_{-2} ]
 \\
\U_{10} &=& \Big(\, L_{-2} L_{-2} L_{-2} L_{-2} L_{-2}
     + { \frac {3821}{225}}\, L_{-4} L_{-2} L_{-2} L_{-2}
 \\ &&\;+ { \frac {657}{50}}\,   L_{-3} L_{-3} L_{-2} L_{-2}
     - { \frac {99}{10}}\,    L_{-4} L_{-3} L_{-3} \,\Big)\,\vec0
 \\[1mm] \hline
\end{array}
}
\]
\vspace{-5mm}
\caption{Conserved currents of the (2/5) and (7/5) perturbations}
\label{tab:1}
\end{table}

\subsection{Zamolodchikov's counting argument}

In \cite{Zamo2b} Zamolodchikov showed how a simple counting argument
can prove the existence of conserved quantities. We recall the method
since some elements of it will be useful later.

Let us consider the simple case of a current $U_{n}(z)$
of conformal weight $n$ which is a polynomial in $L(z)$ and its
derivatives, and a perturbation by a Virasoro primary field
$\phi_h(z)$.
We have
\be
  \left[\, \oint U_{n}\;,\; \oint \phi_h\,\right]
= \oint \psi
\;,
\label{eq:com}
\ee
where $\vec\psi = (U_{n})_{-n+1}\,\vec{\phi_h}$. If the dimension
$d^0_n$ of the space of non-trivial integrals $\oint U_{n}$ is
greater than the dimension $d^h_{n-1}$ of non-derivative
descendants of $\vec\Phi$ at level $n-1$, then the existence of
$d^0_n- d^h_{n-1}$ conserved currents is guaranteed.

If we define the modified character of a Virasoro highest weight
representation of weight $h$ by
\be
  \chi_h = {\mathrm{Tr}}\,q^{h-L_0}
\;,
\ee
then the characters $\tilde\chi_h$ of non-derivative, or
quasi-primary, states are given by
\be
  \tilde\chi_0 = \sum\, d^0_n q^n = (1-q)\,\chi_0 +q
\;,\;\;\;\;
  \tilde\chi_h = \sum\, d^h_n q^n = (1-q)\,\chi_h \;\;\; (h\neq 0,-1\ldots)
\;.
\ee
Applying this to the Potts model, we find (to order $q^{10}$)

\begin{table}[htb]
\[
{ \renewcommand{\arraystretch}{1.4}
\begin{array}{lll}
  \hline
{}
\tilde\chi_0
&=& 1 + q^2 + q^4 + 2\,q^6 + 3\,q^8 + q^9
  + 4\,q^{10} + \ldots
  \\
{}
\tilde\chi_3
&=& 1 + q^2 + q^3 + q^4 + q^5 + 3\,q^6 + 2\,q^7 + 4\,q^8 + 4\,q^9
    + 6\,q^{10} + \ldots~
  \\
{}
\tilde\chi_{2/5}
&=& 1 + q^3 + q^4 + q^5 + 2\,q^6 + 2\,q^7 + 3\,q^8 + 4\,q^9
  + 5\,q^{10} + \ldots
  \\
{}
\tilde\chi_{7/5}
&=& 1 + q^2 + 2\,q^4 + q^5 + 3\,q^6 + 2\,q^7 + 5\,q^8 + 4\,q^9
  + 7\,q^{10} + \ldots
{}
  \\ \hline
\end{array}
}
\]
\vspace{-5mm}
\caption{Characters of quasi-primary states in the Potts model}
\label{tab:2}
\end{table}

Now, examining $\tilde\chi_0 - q\, \tilde\chi_{2/5}$
we can infer the existence of $U_2,U_6,U_8 $
(and, expanding further, $U_{12}$), and examining
$\tilde\chi_0 - q \,\tilde\chi_{7/5}$, of $\U_2,\U_4,\U_6$ and $\U_8$.

This method may be adapted to deduce the existence of the conserved
quantities which are linear in $W$ \cite{Zamo4}.
Since the currents are all Virasoro descendents of $W(z)$, the number
of quasi-primary fields of weight $n$ is given by
$d^3_{n-3}$. Similarly, since the operator product of a Virasoro
descendant of $W(z)$ with $\phi_{2/5}$ is a Virasoro descendant of
$\phi_{7/5}$, the number of quasi-primary fields of weight $2/5 + n-1$
which may occur on the right hand side of (\ref{eq:com}) is given by
$d^{7/5}_{n-2}$. Consequently, to verify the existence of a conserved
current of this form we need to check
$q^3\, \tilde\chi_3 - q^2\,\tilde\chi_{7/5}$, and find that in this
way $U_3,U_5$ and $U_9$ are guaranteed to exist.

\newpage
\section{The general perturbation of the 3-state Potts model}

As seen earlier, the general 3-state Potts chain has 3 parameters
and Cardy proposed that this corresponds to the action
\be
  S_{0} +
\int \!\d^2z\, \left(\,
        \tau\,\Phi_{2/5,2/5} \,+\,
      \delta\,\Phi_{7/5,2/5} \,+\,
  \bar\delta\,\Phi_{2/5,7/5}
           \,\right)
\;.
\label{eq:cact}
\ee
The standard thermal perturbation of the 3-state Potts model is given
by $\delta=\bar\delta=0$ and is integrable with the conserved currents
$U_3,\ldots$.
In \cite{Card3} Cardy showed that the more general model with
$\bar\delta=0$ is also integrable by the following argument. Since the
anti-holomorphic dependence of both fields $\Phi_{2/5,2/5}$ and
$\Phi_{7/5,2/5}$ is the same, i.e. $\bar\phi_{7/5}(\bar z)$, all the
anti-holomorphic conserved currents of the thermal perturbation will
remain  conserved for this double perturbation.
However, a quick glance at table \ref{tab:1} shows that there are no
non-trivial holomorphic conserved currents common to both the
$\phi_{2/5}$ and $\phi_{7/5}$ perturbations, and so it is not clear
what will happen to the holomorphic conserved currents of the thermal
perturbation when the perturbation $\delta$ is turned on.

However, it is important to note that the action (\ref{eq:cact}) no
longer preserves rotational, or Lorentz, invariance, and hence
conserved currents do not have to have a well defined spin. For
example, we can consider currents of the form
\be
  T_{n} =  T_{(n,0)} \;+\; \frac{\delta}{\tau} \,T_{(n,1)}
\;,
\label{eq:f1}
\ee
where $T_{(n,0)}$ and $T_{(n,1)}$ are some conformal fields of weights
$n$ and $n+1$ respectively.
Such a current will be conserved for the doubly perturbed action if
the following three equations~hold:
\bea
&&    \left[\, \oint T_{(n,0)}\,,\, \oint \phi_{2/5} \,\right]
= 0
\;,
\label{eq:a}
\\
&&    \left[\, \oint T_{(n,1)}\,,\, \oint \phi_{7/5} \,\right]
= 0
\;,
\label{eq:b}
\\
&&    \left[\, \oint T_{(n,0)}\,,\, \oint \phi_{7/5} \,\right]
+
    \left[\, \oint T_{(n,1)}\,,\, \oint \phi_{2/5} \,\right]
= 0
\;.
\label{eq:c}
\eea
Eqns. (\ref{eq:a}) and (\ref{eq:b}) imply that
$T_{(n,0)} = \alpha\,U_n$ and $T_{(n,1)}=\beta\,\U_{n+1}$ for some
$\alpha,\beta$.
The extra requirement (\ref{eq:c}) can be ensured by a modification of
Zamolodchikov's argument. In this case,
if there is only a single quasi-primary descendent of $\phi_{2/5}$ at
level $n$ then both terms on the right hand side of (\ref{eq:c}) must
be proportional, and hence cancel for some choice of $\alpha/\beta$.
Examining $\tilde\chi_{2/5}$, we see that this does indeed happen for
$n=3$ and $n=5$. As a result we have proven the existence of
holomorphic conserved currents in Cardy's model.

The next possible  value of $n$ for which there might be a conserved
current of the form (\ref{eq:f1}) is $n=8$, but explicit calculation
shows that this trick will not work.
However, we can instead extend the ansatz to include three terms
\be
  T_{n} = T_{(n,0)} \;+\;
            \frac{\delta}{\tau}\, T_{(n,1)}  \;+\;
             \left(\frac{\delta}{\tau}\right)^{\!2} T_{(n,2)}
\;,
\label{eq:f2}
\ee
where now $T_{(n,0)} = \alpha\,U_n$, $T_{(n,2)}=\beta\,\U_{n+2}$ and we
have the non-trivial requirement that
\be
    \left[\, \oint T_{(n,0)}\,, \oint \phi_{7/5} \,\right]
+
    \left[\, \oint T_{(n,1)}\,, \oint \phi_{2/5} \,\right]
=
    \left[\, \oint T_{(n,1)}\,, \oint \phi_{7/5} \,\right]
+
    \left[\, \oint T_{(n,2)}\,, \oint \phi_{2/5} \,\right]
= 0
\;.
\label{eq:d}
\ee
We have verified that there are conserved currents of this form for
$n=6,8$ by explicit calculation. We include these with the two
previous conserved currents in table \ref{tab:3}, in which we
again give the states $T_{n} = T_{n}(0)\vec 0$ and have set
$\delta/\tau=1$ for simplicity.

\begin{table}[htb]
\[
{\renewcommand{\arraystretch}{1.4}
\begin{array}{lll}
\hline
\qptry2(2) &=& U_2
 \\
\qptry2(3) &=& U_3 - \frac 79 \U_4
 \\
\qptry2(5) &=& U_5 - \frac{91}{180} \U_6
 \\
\qptry2(6) &=& U_6
   - \,\frac{ 147 }{275} \,\vac[
     12 \,L_{-4} W_{-3}
   - 15 \,L_{-3} W_{-4}
   + 10 \,L_{-2} L_{-2} W_{-3} ]\,
   + { \frac {175}{66}}\,\U_8
 \\
\qptry2(8) &=& U_8
   - \,\frac{ 4837476 }{ 1322035 } \,\vac[
     \,L_{-2} L_{-2} L_{-2} W_{-3} \,+\, \ldots\; ]\,
   + { \frac {343}{387}}\,\U_{10}
 \\[1mm] \hline
\end{array}
}
\]
\vspace{-5mm}
\caption{Conserved currents of the (2/5) plus (7/5) perturbation}
\label{tab:3}
\end{table}

Finally we should remark that on dimensional grounds there are no
further corrections to the conservation equation for $T_{n}$ and so
this result should be exact to all orders in perturbation theory.

\section{Remarks and conclusions}

We have shown by counting arguments and explicit calculation that the
double perturbation of the 3-state Potts model considered by Cardy has
extra conserved currents interpolating those known for the two
constituent perturbing fields.

Although we have only constructed four conserved currents, an
appealing pattern has appeared which suggests that there are conserved
currents  $T_{n}$ for all $n=0,2\, \mathrm{mod}\, 3$, polynomial in
$x = \delta/\tau$, and interpolating the conserved currents of the
$\phi_{2,5}$ and $\phi_{7/5}$ perturbations:
\bea
  T_{3n} &=& U_{3n} \,+ \ldots +\,  x^n \,\beta_n\, \U_{4n}
\;,
\nn[2mm]
  T_{3n+2} &=& U_{3n+2} \,+ \ldots +\, x^n \,\beta'_n\, \U_{4n+2}
\;.
\eea

It is interesting to note that the conserved quantities in table
\ref{tab:3} remain formally conserved to first order for the even more
general action
\be
  S_{0} +
\int \!\d^2z\, \left(\,
        \tau\,\Phi_{2/5,2/5} \,+\,
      \delta\,\Phi_{7/5,2/5} \,+\,
  \bar\delta\,\Phi_{2/5,7/5} \,+\,
  \left(\frac{\delta\bar\delta}{\tau}\right)\,\Phi_{7/5,7/5}
           \,\right)
\;.
\label{eq:cact2}
\ee
This again relies on ignoring the $\bar z$ dependence of the
perturbing fields, in which case we can formally factorise the
potential as
\[
  \tau\,\Big(\,\phi_{2/5} \;+\; \frac{\delta}{\tau}\,\phi_{7/5}\,\Big)
      \, \Big(\, \bar\phi_{2/5}
                \;+\; \frac{\bar\delta}{\tau}\, \bar\phi_{7/5} \,\Big)
\;,
\]
and it is clear that the new currents $T_{n}$ are conserved for
(\ref{eq:cact2}), as are similar currents $\bar T_{n}$ constructed
{}from the anti-holomorphic algebra.
This is not sufficient to conclude that this potential is
integrable -- the first order in perturbation theory is no longer
exact since $( \delta^3 \bar \delta^3 /\tau )$ is dimensionless and
there are possible corrections to the conservation equation at
arbitrarily high orders in perturbation theory. Furthermore, the
explicit presence of an irrelevant field in the action spoils the
property that the UV limit of the perturbed model is simply the
conformal field theory.

An interesting point to notice is that the results of \cite{AMco1}
indicate that $\phi=\varphi=\pi/2, 0.901.. < \lambda < 1.1095.. $ has
massless modes described by a
parity violating theory with $c=\bar c=1$.
It is believed that these values are
continuously connected to the conformal point $\phi=\varphi=0,\lambda=1$
through massless theories, but it is hard to see how they can be
reached by perturbation of the conformal 3-state Potts, since the
central charge of the conformal 3-state Potts model is $4/5$ and one
might expect central charge to decrease along renormalisation group
flows. Perhaps the presence of an irrelevant field in the perturbation
signals that it is a perturbation from a model with larger $c$, as
happens e.g. for the Virasoro minimal models where the irrelevant
$\phi_{3,1}$ perturbation of the $M_p$ model corresponds to the IR
limit of the $\phi_{1,3}$ perturbation of the $M_{p+1}$
model. However, one should remember that the true state of lowest
energy having non-zero momentum and that it may be very hard to relate
the exact results to those obtained in the perturbed conformal models.

It is also interesting to note that Cardy finds a different
two-dimensional subspace of the space of coupling constants to be
integrable ($\bar\delta=0$) to that which appears to be the case from
the Transfer matrix approach ($\tau\sim\delta\bar\delta$) suggesting
that it might be possible that both results are correct, consistent
with the action (\ref{eq:cact2}) being integrable for all values of
the coupling constants. Although the spin chain is not believed to be
integrable for all values \cite{MOrr1}, it is possible that the
scaling limit of the spin chain only differs from an integrable model
by irrelevant operators, which, while breaking the exact integrability
of the spin chain would give an integrable model in the IR. However,
as Cardy notes, it is also possible that there is no relation between
lattice integrability and the integrability of perturbed conformal
models.

We should like to mention that there are well-known models which
contain dimensionless parameters and which are believed to be
integrable, for example the staircase models \cite{stair}. These are
double perturbations of a conformal field theory by a relevant and an
irrelevant operator which share the same conserved currents (to first
order). While it is not possible to check integrability by exhibiting
conserved currents exactly for the reason that there are dimensionless
parameters, they do appear to share many properties of integrable models.

Finally, we should like to discuss possible generalisations of these
results.
Cardy's argument is sufficient to show that given any W-algebra and an
integrable perturbation $\Phi$ then the anti-holomorphic conserved
quantities for $\Phi$ remain conserved for any perturbation of the
form $W_{-1}\Phi$. By contrast, to be able to apply our method to
find holomorphic conserved currents for such a perturbation, it is also
necessary that $ W_{-1}\Phi$ is an integrable perturbation for some
subalgebra of the W-algebra. In the model treated the original
perturbation is $\Phi_{2/5}$ and $W_{-1}\Phi_{2/5}$ is an integrable
perturbation for the  Virasoro subalgebra of the $W_3$
algebra. However, it is easy to check that there are no other rational
models of the $W_3$ algebra for which $\Phi$ is an integrable
perturbation and $W_{-1}\Phi$ is an integrable perturbation of the
Virasoro subalgebra. 

The next obvious possible generalisation is the $Z_n$ chiral Potts
models. These are described by a spin chain Hamiltonian
given in terms of $(n\times n)$ matrices $\sigma,\Gamma$, and again
dependent on three parameters $\lambda,\phi,\varphi$.
The point $\lambda=1,\phi=\varphi=0$ is now described by a
$c=c_n = 2(n-1)/(n+2)$ conformal field theory which can be variously
understood as the $Z_n$ parafermion model \cite{FZam2b}, the first
unitary minimal model of the $W_n$ algebra \cite{FLuk1} or a model
with a $W(2,3,4,5)$ chiral algebra \cite{unif}. The thermal
perturbation is given by a field of weight $h=h_n=2/(n+2)$ and Cardy has
proposed that the chiral perturbation corresponds to a linear
combination of level~1 W-descendents of this field,
$\alpha_j W^{(j)}_{-1}\vec h$. Given the results above, it is natural
to suppose that this is itself an integrable perturbation for the
subalgebra of the W-algebra which is invariant under the automorphism
which sends the odd-spin generators $W \to -W$. The full W-algebra and
its orbifold have been studied in some detail \cite{DBFH1,unif}, and
it is believed that for $c=c_n$ the orbifold algebra is of
the form $W(2,4,6,8)$. While it is not yet possible to study the
representations of this algebra directly, there is some evidence that
it is in turn a truncation of the $\WB_{(n-1)/2}$ algebra%
\footnote{Clearly we restrict to $n$ odd in the following discussion.}
at $c=c_n$ -- the self-coupling of the spin 4 field is the
same in both cases \cite{Horn2,unif}.

It now remains to examine the representations of the $\WB_{(n-1)/2}$
algebras at the $c$-values $c_n$. From \cite{FKWa1} we see that
this is a minimal model of $\WB_{(n-1)/2}$ with primary fields of
weights $h_{\lambda,\mu}$ indexed by $\lambda$, an integrable weight
of $B_{(n-1)/2}$ of level $2$, and $\mu$, an integrable weight of
$C_{(n-1)/2}$ of level $3$. In particular we find amongst the allowed
representations the values
$ h_{0,\Lambda_1^\vee} = 2/(n+2) = h_n$,
$ h_{0,2\Lambda_1^\vee} = (n+4)/(n+2) = h_n+1$ and
$ h_{0,3\Lambda_1^\vee} = 3$, where $\Lambda_1^\vee$ is the first
fundamental weight of $C_{(n-1)/2}$.

These three equations suggest strongly that
for $c=c_n$ the $\WB_{(n-1)/2}$ algebra can be augmented by a
representation of weight 3 to give the full $W_n$ algebra;
that the thermal perturbation is given by the field
$(0,\Lambda_1^\vee)$, which is an integrable perturbation
for the $\WB_{(n-1)/2}$ algebra corresponding to the ATFT
$\smash{a_{n-1}^{(2)}}$;
and that there is a $W_n$ descendent of this field at level 1 which is
itself a highest weight of the $\WB_{(n-1)/2}$ algebra of weight
$h_n + 1$ of type $(0,2\Lambda_1^\vee)$. Since this corresponds
to the ATFT \cnm1, it is also an
integrable perturbation for $\WB_{(n-1)/2}$.

Thus we suggest that the whole procedure in this article will also carry
through for the $Z_n$ chiral Potts models. The thermal perturbation
corresponds to the $\smash{a_{n-1}^{(1)}}$ ATFT with conserved currents of
spins $1,\ldots,n-1$ mod $(n-1)$, and the descendent at level 1 to
the \cnm1\ ATFT with conserved currents of all even
spins, and we expect conserved currents for the double
perturbation interpolating these.

\section{Acknowledgements}

I would like to thank
A. Honecker for many useful discussions on the chiral Potts model and
the orbifolds of $W$--algebras, and P.E.~Dorey, W.~Eholzer and R.A.\
Weston for useful discussions, and the organisers of the 1996 E\"otv\"os
summer school on CFT and integrable models in Bolyai College,
Budapest, for their hospitality while this work was started.

GMTW is grateful to the EPSRC for an advanced fellowship and the
support of grant GR/K30667.
The calculations in this article were carried out in Maple using a 
package for conformal field theory calculations written by H.~G.~Kausch.

\newpage
\small\raggedright

\end{document}